\documentstyle[epsf]{article}

\begin{document}
\setlength{\baselineskip}{0.27in}

\newcommand{\beq}{\begin{equation}}
\newcommand{\eeq}{\end{equation}}
\newcommand{\beqa}{\begin{eqnarray}}
\newcommand{\eeqa}{\end{eqnarray}}
\newcommand{\lsim}{\begin{array}{c}\,\sim\vspace{-21pt}\\<
\end{array}}
\newcommand{\bi}{\bibitem}
\newcommand{\gsim}{\begin{array}{c}\sim\vspace{-21pt}\\>
\end{array}}

\begin{center}
\vglue .06in
{\Large \bf  What Do We Know About the Tau Neutrino?}
\\[.5in]
\begin{tabular}{c}
{\bf  I.Z. Rothstein}\\[.05in]
{\it The Randall Laboratory of Physics}\\
{\it University of Michigan}\\
{\it Ann Arbor MI 48109 }\\[.15in]
\end{tabular}
\vskip 0.25cm
{\bf Abstract}\\[-0.05in]

\begin{quote}
In this talk I review the present bounds on the tau neutrino, concentrating on
the
possibility of bringing the mass bound  below an MeV.
\end{quote}
\end{center}
\section{Introduction}
Given the recent evidence for the detection of the top quark, all the
particles in the standard model have been directly detected, save for
the tau neutrino and the Higgs particle. Of course, while it would not
be difficult to believe in a world with no Higgs, one would be rather
hard pressed to find a theory consistent with the data that did not
contain a tau neutrino.  Thus, I mention the direct detection issue
only to point out that, compared to the other leptons, we know very
little about the tau neutrino.

Neutrinos are notoriously slippery objects as a consequence of their
small masses and cross sections. One need only look at the checkered
history of neutrino mass searches to appreciate this fact. The tau
neutrino is even more troublesome to detect, compared to the other
neutrino species, due the short lifetime of the tau lepton. Below I
will review the experimental status of tau neutrino mass searches as
well as bounds on the tau neutrino magnetic moment.  I will deal here only
with direct searches.
As will be seen,
the bounds are rather weak compared to the other neutrino species.

More stringent limits can be obtained from the fact the the primordial helium
abundance is sensitive to massive neutrinos. This fact is really a
fortunate numerical accident. The present lab bounds on the tau
neutrino mass\cite{aleph} (24 MeV) is just in the range where
primordial nucleosynthesis feels the effects of the mass.  The
nucleosynthesis bounds will rule out neutrino masses from the lab
bound down to a few hundred keV. The great success of the primordial
nucleosynthesis predictions give us confidence in these bounds. However, as
Gary Steigman discussed in his talk, there have been some
difficult issues raised lately regarding a dearth of Helium. I will
not discuss this issue here and refer the interested  reader to Steigmans' talk
in
these proceedings.

The bounds derived from primordial nucleosynthesis will only apply for
lifetimes greater than $O(100)~sec.$ Thus, if we really want to rule
out MeV neutrinos, then a detailed study of possible decay modes is
necessary. In this talk I will address the issue of excluding possible
decay windows. Some decay modes may be ruled out without recourse to
model considerations. But for  other modes  I will be forced to resort to
theoretical arguments to rule out regions of parameter space.

It will be shown that an MeV neutrinos will most probably have to decay rapidly
($\tau<O(100)~sec.$) to a
massless scalar or three light neutrinos. Cosmologically, a  tau neutrino mass
may fit quite nicely.  Given the family hierarchy
structure it is expected that the tau neutrino should be the heaviest
of the three neutrinos. Thus, it is  tempting to use a heavy tau
neutrino to address several cosmological issues. For instance, an MeV
Majorana neutrino mass could play a role in generating a lepton number
in the early universe, which in turn could be transformed into a net
baryon asymmetry through non-perturbative effects \cite{krs}.   Thus, ruling
out MeV
neutrinos would exclude a body of theoretical proposals.

\section{Lab Bounds}
Recently the lab bound on the tau neutrino mass has decreased from 30
MeV to 24 MeV at $95\%$ CL. This bound comes from looking for missing
energy in the decay $\tau \rightarrow 5\pi^{\pm}(\pi^0) $.  The
improvement in the bound was rather fortuitous as it essentially comes
from one event.  Prospects for improving the bound are limited by the
transverse momentum resolution.

The bounds on the magnetic moment are:
\begin{eqnarray}
&   \mu_{\tau \tau}&<5.4\times 10^{-7}\mu_B~^{3}\nonumber \\
&   \mu_{\tau ?}&<10^{-9}\left( {m_\nu\over{MeV}}\right)^2\mu_B~^{4}\nonumber
\\
&   \mu_{\tau ?}&<4.0 \times 10^{-6} \mu_B ~^{5}.
\end{eqnarray}
 If there is no mass dependence stated above, then the bound is mass
independent. I've copied these bounds here, because, as we will see
later, ruling out MeV neutrinos will entail the use of these
bounds\footnote{There are more stringent constraints on the magnetic
moment coming from stellar cooling. But they will not apply for MeV
neutrinos.}.  The bounds for the mass and the magnetic moments, are
much less stringent than the bounds for the lighter neutrinos.
However, we may do much better using constraints from cosmology and
astrophysics.

\section{Cosmological and Astrophysical Bounds}
It has been known for quite some time that  a neutrino species with mass in the
range
\cite{gz}\cite{hut}
\begin{equation}
90~eV \lsim m_\nu \lsim~ 2 ~GeV
\end{equation}
would lead to our universe being much younger than it is. The reason for this
is
that as we increase the energy density in neutrinos, the expansion
rate of the universe increases, thus the universe would have evolved
to its present state is a much shorter period. This is a statement of
flatness. Given the smallness of the spatial curvature, we see that
the potential energy is commensurate with the kinetic, thus any
increase in the potential energy must result in a faster expansion
rate.

While this bound is quite stringent, it is only valid for neutrinos which are
effectively
stable, and, as such not as powerful as we would have hoped. There are several
ways in
which one can do better. Let us first consider the effects of an MeV
neutrino on the primordial element abundances.  The relevant scale for
nucleosynthesis is $O(MeV)$, thus we would expect that a neutrino species with
a
mass  in the range we're interested in could indeed have an effect (this is
the fortuitous numerics mentioned in the introduction).  A neutrino species
with an MeV mass will
enhance the energy density relative to the contribution from a
massless neutrino at 1 MeV\cite{ks}\cite{dr}\cite{sky}. This is because the
energy density in the
massless species will be redshifted and hence diluted relative to the
 energy density of  a massive species.
The resulting increased expansion rate will have the effect of
increasing the neutron to proton ratio at the time of the freeze out
of the weak interactions, which is at about $T \approx 1 MeV$.  Once
the weak interaction freezes out the neutron to proton ratio will
decrease only as a result of free neutron decay.  This decrease will
continue until the onset of deuterium formation, which begins at
$T=.065~MeV$.  Essentially all the free neutrons left at this
temperature will end up in Helium.

One calculates the final Helium abundance by solving the Boltzmann
kinetic equation, as has become routine for practitioners in the
field\cite{gg}.  Here I only mention an effect that differs from the standard
calculations.  Since our result will be sensitive to the  time between weak
freeze out and  deuterium formation we must be mindful to calculate
the time temperature relation with care. In the case where the energy
density is dominated by either relativistic or non-relativistic
species the time temperature relationship is given by
\begin{equation}
{d\over{dt}}T=-H T.
\end{equation}
However, if there are relativistic and non-relativistic species
contributing nearly commensurate amounts to the energy density, and
furthermore they are exchanging energy, then covariant energy
conservation demands that the time temperature relation be \cite{dr}
\begin{equation}
{d\over{dt}}T=-H\left(1+{0.2 r \over{g^{rel}_*+0.42 r{m\over{T}}}}
\right)
-{.14 r({3\over{2}}+{m\over{T}})\over{g^{rel}_*+0.42 r{m\over{T}}}}.
\end{equation}
Where $g_*^{rel} $ counts the number of degrees  of freedom
in relativistic species and r is the ratio of the number density of
massive neutrinos to massless.

After solving the kinetic equations necessary to calculate the Helium
abundance it is found that for majorana masses the forbidden mass
range is \cite{dr}
\begin{equation}
0.5<m_m<35 MeV,
\end{equation}
and for Dirac neutrinos the forbidden range is
\begin{equation}
.3<m_D<35~MeV.
\end{equation}
This range is found by imposing the constraint that the net Helium
abundance be less than the amount of Helium that would have been
produced had there been 3.3 massless Weyl neutrino species.  I refer
the reader to Gary Steigmans' talk in these proceedings for
considerations regarding the confidence of this bound.

In arriving at the bounds on the Dirac mass we have assumed that the
right handed species does not reach thermal equilibrium below the
temperature of the QCD phase transition, if the neutrino mass is less
than 300 keV\cite{fm}. However, we can do better than this\cite{dkr}. Even if
the right handed neutrino does not reach thermal equilibrium, its out
of equilibrium production rate can yield a contribution to the energy density
which can
still enhance the Helium abundance. Right handed neutrinos can be
produced through the following  processes
\begin{eqnarray}
\pi^0 &\rightarrow& \nu_{\tau+}
{\bar \nu}_{\tau+}  \nonumber \\
l_1l_2& \rightarrow & l_3 \nu_{\mu(\tau)+}\\
l{\bar l}& \rightarrow &\nu_{\mu(\tau)+}{\bar \nu_{\mu(\tau)+}}.
\nonumber
\end{eqnarray}
In this list the $+$ subscript refers to the ``wrong'' helicity state,
that is, right handed neutrinos and left handed anti-neutrinos.  Once
we take into account these out of equilibrium processes it is found
that the upper bound on the Dirac neutrino mass is 190 keV if the
temperature of the QCD phase transition is assumed to be 200 MeV. We
again use the constraint $N_{eff}<$3.3. The bound depends upon the
temperature of the QCD phase transition because all right handed
neutrinos produced prior to the transition will be have their energy diluted
due to the
net entropy dumped into the bath from the transition. Thus as the
temperature of the transition is raised the bound becomes more
stringent.

The bounds discussed above all assumed that the neutrino is stable
on the time scale of nucleosynthesis, which is about 100 seconds. What
if the neutrino decays with a lifetime shorter than this scale?  Is
this a viable scenario?  Consider the possible decay modes:

\begin{eqnarray}
(i)~~\nu_\tau &\rightarrow&3 \nu \nonumber \\
(ii)~~\nu_\tau &\rightarrow&\nu+\gamma ~or~\nu e^+e^-\nonumber \\
(iii)~~\nu_\tau& \rightarrow & \nu+H.
\end{eqnarray}

We would like to be able to eliminate possibilities in a model
independent fashion.  However, for the decay into three neutrinos the
observational consequences  for laboratory experiments
are  nil.  There is one interesting effect  for such decays. It has been
pointed out in ref.\cite{ts},
that an MeV neutrino decaying into electron neutrinos can
affect  nucleosynthesis yields  by distorting the electron neutrino phase space
distribution.  Such distortions can lead to either increasing or decreasing the
Helium abundance depending on the mass. For masses less than 10 MeV,
the Helium abundance can be drastically reduced\cite{dgt}.

Let us consider rapid three neutrino decay from a model building standpoint.
The trouble  with this decay  mode is that it necessitates flavor changing
neutral
currents(FCNC). In general  it is  not a problem to generate FCNC's. The
problems
arise when one tries to generate large FCNC's  in the leptonic sector while
keeping the FCNC's small in the quark sector.  This problem can be avoided
by disentangling the sectors via the imposition of a symmetry, usually in the
Higgs
sector of the theory. This will, in general, necessitate the introduction of
more
Higgs multiplets. One possibility is to mediate the decay through a scalar
triplet (see figure 1a).
As long as  triplet doesn't carry baryon number it  will couple only to the
leptons.  Given that there will  be no GIM mechanism in action due to the fact
that the neutrinos will get masses
from  other scalars (the left handed triplet vev is constrained by the rho
parameter), we can have large FCNC's.  Such a scenario fits into a
left right symmetric model \cite{rs}.  There is one caveat however,  the
charged
scalar in the triplet will mediate the decay $\nu_\tau\rightarrow  e^+e^-
\nu_e$, as shown in figure
1b.  Such a  decay would distort the light curves for SN1987A, and as such are
ruled out
\cite{fs}.  One could however, avoid this dilemma by having the tau neutrino
decay
into  $\nu_\mu +\nu_e \bar{\nu_e}$.   As was pointed out in \cite{mn}, this
mode will not be
constrained by exotic tau decay bounds, as one might naively expect,  unless
the bound
on the leptonic KM angle $\theta_{\tau \mu}$ is improved by an order of
magnitude.
Thus, three neutrino decay is viable in this scheme as long as we are willing
to supress certain Yukawa couplings.  There may be other viable schemes as
well, but it
seems clear that  any such scheme will demand physics quite a bit beyond the
standard model.

 We may virtually eliminate the possibility of decay (ii) without
recourse to model building considerations.  Due to the luminous nature
of the decay products there are many observations which disallow this
decay mode. I will not reiterate here all the arguments and instead
refer the reader to \cite{bgr} for the details. Figure 1. shows the
mass/lifetime exclusion plot. These bounds come from combining the
bounds from eq(1) with those from supernova considerations. Notice
that there is a small window for extremely rapid decay
$(\tau<10^{-12}sec.)$ into a sterile neutrino and photon. Thus, it is
not possible to completely rule out the possibility for this decay,
however, such a rapid decay seems very implausible. Furthermore, I
believe that a detailed nucleosynthesis calculation would close this
tiny window.
\begin{figure}
\centering
\epsfysize=3in   
\hspace*{0in}
\epsffile{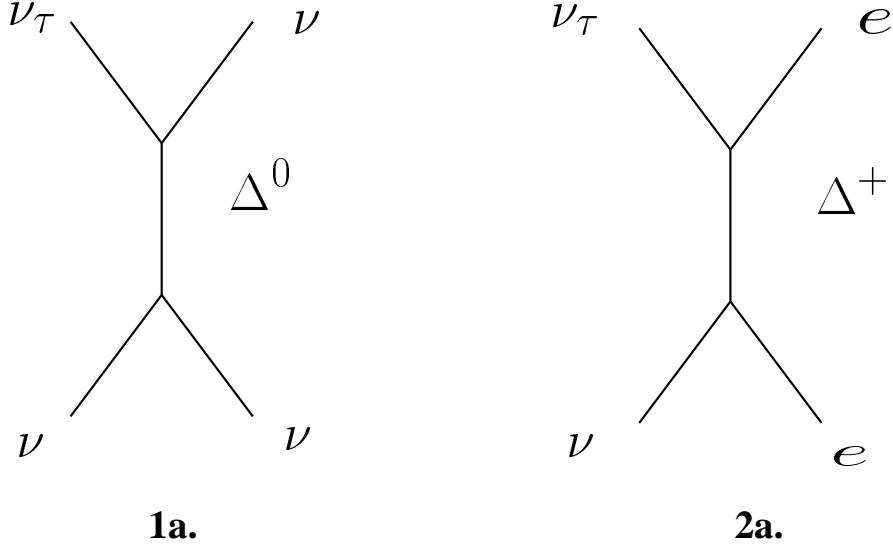}  
\caption{1a.Rapid  neutrino decay through triplet exchange. 1b. Contribution
to  disallowed $\nu_\tau$ decay   through the charged partner.}
\label{leadjet}  
\end{figure}

Finally let us consider the last decay mode (iii). This mode has been
looked at carefully in \cite{sky}. The authors found that this is
indeed a viable mode, and that this decay is allowed for lifetimes
less than 40 seconds.  There are some ranges of masses and lifetimes
which satisfy this constraint which are disallowed due to the fact
that the scalar can contribute significantly to the energy density.
{}From a model building perspective this decay is perhaps easier
to implement than modes (i)  and (ii). A simple extension of the standard model
including just
a singlet Higgs and  spontaneous lepton number violation leads
to neutrino decay into  a massless Goldstone boson (Majoron) \cite{cmp}.
It was originally believed that such a rapid decay is not viable in  the
simplest version
of the Majoron model
\cite{vs}.   However, as has been  pointed out,  this is not  necessarily the
case. Rapid decays are feasible
as long as there  exists  a hierarchy in the dirac masses \cite{jl}.
\begin{figure}
\centering
\epsfysize=5in   
\hspace*{0in}
\epsffile{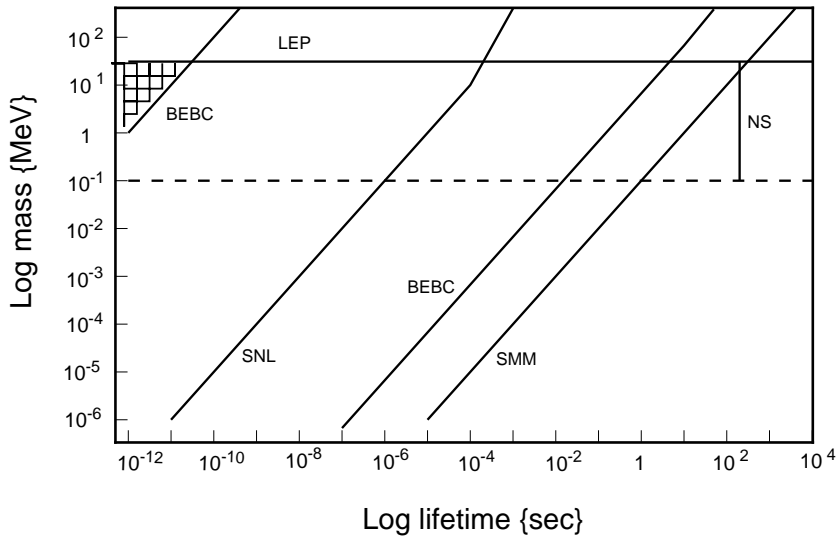}  
\caption{ Bounds on radiative and $e^+e^-$ decays of $\nu_\tau$.
Curve labels lie on the forbidden side of curves: lab mass bound (LEP),
nucleosynthesis (NS),
supernova luminosity (SNL)\protect \cite{fs}, Solar Max Mission (SMM)\protect
\cite{kt}.  The hatched region is allowed
for decays into sterile neutrinos. The decay in active neutrinos is disallowed
by bounds on off diagonal  moments of  active neutrinos. }
\end{figure}

Before closing let us return to the case of Dirac masses. It is
possible to get a more stringent bound than that obtained from
nucleosynthesis considerations from supernova luminosity
arguments. As was pointed out in \cite{t1}, if the Dirac mass is
greater than 15 keV and its lifetime is greater than $10^{-6}$
sec$(m/MeV)$, then the right handed species would have been generated
in the core of SN1987A .  These right handed neutrinos, if sterile
with respect to the composition of the supernova, would have rapidly
depleted the core energy.  Such an energy depletion would have
shortened the length of the observed neutrino pulse.  If the neutrino
decays into a electron or muon neutrinos, then the more stringent
constraint of 1 keV \cite{t2} may be obtained by noticing that the
neutrinos emitted from the core without thermalizing would be much
more energetic than the observed neutrinos.

While these bounds on Dirac masses from supernova arguments are more
stringent, I point out that they have the drawback that they assume
that the right handed species is sterile \cite{bmr1}. As such, these
are model dependent. While it is true that if
right handed neutrinos do exist it would seem reasonable that they
would be sterile, we must not jump to any conclusions.

\section{Conclusions }
Given the arguments discussed above what can we say we know about the
tau neutrino?  Aside from the statement that it must be there, we can
only say that its mass should be less than 400 keV if has a Majorana
mass, and less than 200 keV if has a Dirac mass. These bounds can be
avoided if it decays into a massless scalar and a light neutrino, or three
neutrinos with
a lifetime less than  O(100) seconds. If we are willing to assume that the
right handed species is sterile then we may rule out Dirac masses
greater than 15 keV for lifetimes greater than $10^{-6}$ sec$(m/MeV)$.
Thus, if the tau neutrino is found to have masses in the disallowed
ranges, then it would mean that we would need to invoke physics quite far
beyond
the standard model.  The more  likely scenario is  that the tau
neutrino, if massive, will have masses below the MeV range.

What hope do we have of lowering the bound on the tau neutrino mass? Or for
that
matter, learning anything more about the tau neutrino?   The best prospect for
lowering
the lab bound on the mass is at the B factory, where  a bound of a few MeV can
be reached \cite{gc}.
It seems that  to do better than this we must use
indirect meaurements of the mass through oscillations experiments.

\end{document}